\documentclass[12pt]{iopart}
\usepackage{graphicx}

\begin{document}
\def \ee {\varepsilon}
\thispagestyle{empty}
\title[]{On the definition of dielectric permittivity
for media with temporal dispersion in the presence
of free charge carriers
}

\author{
M.~Bordag,${}^{1}$
B.~Geyer,${}^{1}$
G.~L.~Klimchitskaya${}^{1,2}$
and V.~M.~Mostepanenko${}^{1,3}$
}

\address{${}^1$Institute for Theoretical
Physics, Leipzig University, Postfach 100920,
D-04009, Leipzig, Germany}

\address{$^2$North-West Technical University, Millionnaya St. 5,
St.Petersburg, 191065, Russia
}

\address{$^3$
Noncommercial Partnership  ``Scientific Instruments'',
Tverskaya St. 11, Moscow, 103905, Russia}

\begin{abstract}
We show that in the presence of free charge carriers the
definition of the
frequency-dependent dielectric permittivity requires additional
regularization.
As an example, the dielectric permittivity of the Drude
model is considered and its time-dependent counterpart is derived and
analyzed. The respective electric displacement cannot be represented
in terms of the standard Fourier integral.
The regularization procedure allowing to circumvent these difficulties
is suggested.
For purpose of comparison
it is shown that the frequency-dependent dielectric permittivity
of insulators satisfies all rigorous mathematical criteria.
This permits us to conclude that in the presence of free charge
carriers the concept of dielectric permittivity is not as well defined as for
insulators and we make a link to widely discussed puzzles in the theory
of thermal Casimir force which might be caused by the use of such
kind permittivities.
\end{abstract}

\section{Introduction}

The concept of dielectric permittivity in media with temporal
dispersion is commonly used in electrodynamics and condensed matter
physics (see, e.g., \cite{1,2}). For not too strong fields the
dielectric permittivity $\varepsilon(\tau)$ depending on the
time-like variable $\tau$ is introduced from the linear integral
relation between the electric field
$\mbox{\boldmath$E$}(\mbox{\boldmath$r$},t)$ and the electric displacement
$\mbox{\boldmath$D$}(\mbox{\boldmath$r$},t)$.
Then the frequency-dependent permittivity $\varepsilon(\omega)$ is
defined using the Fourier transformations  of the fields
$\mbox{\boldmath$E$}(\mbox{\boldmath$r$},t)$ and
$\mbox{\boldmath$D$}(\mbox{\boldmath$r$},t)$.
Below we argue that this procedure, which is wholly satisfactory for
dielectric materials, faces additional
regularization problems in an infinite medium containing free
charge carriers. This leads us to the conclusion that some applications
of the frequency-dependent dielectric permittivities allowing for free
charge carriers might be not rigorously justified. As one such example we
discuss the dielectric permittivity of the Drude model which leads
to widely discussed difficulties when substituted into the Lifshitz
formula for the van der Waals and Casimir force at nonzero temperature
\cite{3}.

\section{Dielectric permittivity in the presence of temporal
dispersion}

For simplicity we consider an isotropic nonmagnetic medium
of infinite extent. If its
properties do not depend on time, the linear dependence between
$\mbox{\boldmath$D$}(t)$  and
$\mbox{\boldmath$E$}(t)$ satisfying causality
(here and below we omit the argument $\mbox{\boldmath$r$}$)
is given by
\begin{equation}
\mbox{\boldmath$D$}(t)=\int_{-\infty}^{t}dt^{\prime}\,
\varepsilon(t-t^{\prime})\mbox{\boldmath$E$}(t^{\prime}).
\label{eq1}
\end{equation}
\noindent
The kernel $\varepsilon(t-t^{\prime})$ of the integral operator
on the right-hand side of
(\ref{eq1}) is called the dielectric permittivity for media with
temporal dispersion.
We represent it in the form
\begin{equation}
\varepsilon(t-t^{\prime})=2\delta(t-t^{\prime})+f(t-t^{\prime}),
\label{eq2}
\end{equation}
\noindent
where $f(t-t^{\prime})$ is a continuous real-valued function  and
the delta function $\delta(t-t^{\prime})$ is defined on the interval
$-\infty < t^{\prime}\leq t$ in the following manner
\cite{Korn}:
\begin{equation}
\int_{-\infty}^{t}g(t^{\prime})\,\delta(T-t^{\prime})\,dt^{\prime}
=\left\{
\begin{array}{l}
0,{\ \ }T>t, \\
\frac{1}{2}g(T-0),{\ \ }T=t, \\
\frac{1}{2}\bigl[g(T-0)+g(T+0)\bigr],{\ \ }-\infty<T<t.
\end{array}
\right.
\label{eq2a}
\end{equation}
\noindent
Here, $g(t)$ is an arbitrary function which has  bounded variation
in the vicinity of the point $t^{\prime}=T$.
Note that from physical point of view
the function $f(t-t^{\prime})$ is defined only for
$t^{\prime}\leq t$; for $t^{\prime}>t$ it can be ascribed any values
(including to vanish in that region).

Substituting (\ref{eq2}) into (\ref{eq1}) with account of (\ref{eq2a})
and introducing the new variable $\tau=t-t^{\prime}\geq 0$,
we rearrange (\ref{eq1}) to \cite{1}
\begin{equation}
\hspace*{-7mm}
\mbox{\boldmath$D$}(t)=\mbox{\boldmath$E$}(t)+
\int_{-\infty}^{t}dt^{\prime}\,
f(t-t^{\prime})\mbox{\boldmath$E$}(t^{\prime})
=
\mbox{\boldmath$E$}(t)+
\int_{0}^{\infty}d\tau\,f(\tau)\mbox{\boldmath$E$}(t-\tau).
\label{eq3}
\end{equation}
\noindent
Representing the real functions $\mbox{\boldmath$D$}(t)$  and
$\mbox{\boldmath$E$}(t)$ as Fourier integrals,
\begin{equation}
\mbox{\boldmath$D$}(t)=\int_{-\infty}^{\infty}
\mbox{\boldmath$D$}(\omega){\rm e}^{-{\rm i}\,\omega t}d\omega,\qquad
\mbox{\boldmath$E$}(t)=\int_{-\infty}^{\infty}
\mbox{\boldmath$E$}(\omega){\rm e}^{-{\rm i}\,\omega t}d\omega,
\label{eq4}
\end{equation}
\noindent
one can rewrite (\ref{eq3}) in terms of Fourier transforms
of the fields \cite{1}
\begin{eqnarray}
\hspace*{-7mm}
\mbox{\boldmath$D$}(\omega)=\varepsilon(\omega)
\mbox{\boldmath$E$}(\omega), \qquad
\varepsilon(\omega)\equiv 1+\int_{0}^{\infty}\!\!d\tau\,
f(\tau)\,{\rm e}^{{\rm i}\,\omega\tau}
=\int_{0}^{\infty}\!\!d\tau\,
\varepsilon(\tau)\,{\rm e}^{{\rm i}\,\omega\tau},
\label{eq5}
\end{eqnarray}
\noindent
where $\mbox{\boldmath$D$}(\omega)$,
$\mbox{\boldmath$E$}(\omega)$ and
$\varepsilon(\omega)$ are complex-valued functions.
{}From (\ref{eq5}) it follows that $\varepsilon(\omega)$ is
an analytic function in the upper half-plane of complex $\omega$
including the real axis with possible exception of the point
$\omega=0$. As a result, the real and imaginary parts of
$\varepsilon(\omega)$ are connected by means of the
Kramers-Kronig relations \cite{1}.
Note that in contrast, for instance, to \cite{9} we always
consider fields defined in $(\mbox{\boldmath$r$},t)$-space
as real and only their Fourier transforms might be complex.

The equivalence between (\ref{eq3}) and (\ref{eq5}) requires
the existence of integrals (\ref{eq4}) (and respective inverse
Fourier transformations) and the possibility to change the order
of integrations with respect to $dt^{\prime}$ and $d\omega$.
In mathematics there are many different conditions  on how
to assign a rigorous meaning  to (\ref{eq4}) and respective inverse
formulas. The most widely used demand is that the function
$\mbox{\boldmath$D$}(t)$ should have a bounded variation and be integrable
together with its modulus, i.e., should belong to $L^{1}(-\infty,\infty)$.
In this case the function
$\mbox{\boldmath$D$}(\omega)$ is also bounded, uniformly continuous
on the axis $(-\infty,\infty)$ and $\mbox{\boldmath$D$}(\omega)\to 0$
when $|\omega|\to\infty$ \cite{4}. The function $\mbox{\boldmath$E$}(t)$
should possess the same properties. The change of order of integrations
is possible if both integrals under consideration are uniformly
convergent.

\section{Media with free charge carriers}

It can be easily seen that the above conditions
permitting to introduce the frequency-dependent dielectric
permittivity $\varepsilon(\omega)$ in accordance with (\ref{eq5})
are not directly applicable for
media with free charge carriers. As an example we consider the widely
used dielectric permittivity of the Drude model,
describing such media \cite{5},
\begin{equation}
\varepsilon_D(\omega)=1-\frac{\omega_p^2}{\omega(\omega+{\rm i}\,\gamma)},
\label{eq6}
\end{equation}
\noindent
where $\omega_p$ is the plasma frequency and $\gamma>0$ is the relaxation
parameter.
It is obvious that $\omega=0$ may lead to mathematical problems in
Fourier transformation.
To make a link between $\varepsilon_D(\omega)$ and real-valued
physical fields $\mbox{\boldmath$E$}(t)$ and $\mbox{\boldmath$D$}(t)$,
it would be of interest to determine the respective function
$f_D(\tau)$. The substitution of (\ref{eq6}) into (\ref{eq5}) leads
to the following equations
\begin{eqnarray}
-\frac{\omega_p^2}{\omega^2+\gamma^2}&=&\int_{0}^{\infty}
f_D(\tau)\,\cos(\omega\tau)\,d\tau,
\nonumber \\
\frac{\omega_p^2\gamma}{\omega(\omega^2+\gamma^2)}&=&\int_{0}^{\infty}
f_D(\tau)\,\sin(\omega\tau)\,d\tau.
\label{eq7}
\end{eqnarray}
\noindent
{}From the first equation, by means of the inverse cosine Fourier
transformation performed with the help of the integral 3.723(2)
in \cite{6}, one finds
\begin{equation}
f^{\rm cos}_D(\tau)=-\frac{\omega_p^2}{\gamma}\,{\rm e}^{-\gamma\tau}.
\label{eq8}
\end{equation}
\noindent
It is easily seen that the substitution of (\ref{eq8}) into the first
equation of (\ref{eq7}) with
account of 3.893(2) in \cite{6} leads to a correct identity,
however, substituted into the second equation of (\ref{eq7}) fails.
On the other hand,
using the inverse sine Fourier transformation and the integral
3.725(1) in \cite{6}, the second equation of (\ref{eq7}) leads
to a different result
\begin{equation}
f^{\rm sin}_D(\tau)=\frac{\omega_p^2}{\gamma}\left(1-
{\rm e}^{-\gamma\tau}\right). 
\label{eq9}
\end{equation}
\noindent
Now, the substitution of (\ref{eq9}) into the right-hand sides of
equations (\ref{eq7}) reproduces their left-hand sides up to additional undefined terms and
thus also violates the equalities.

The pathological properties under
consideration are explained by the fact that $\varepsilon_D(\omega)$
results in $\mbox{\boldmath$D$}(\omega)$ which is unbounded in any
vicinity of $\omega=0$. This means that $\mbox{\boldmath$D$}(t)$
cannot be represented as a Fourier integral (\ref{eq4}) and both the
definition of $\varepsilon(\omega)$ in (\ref{eq5}) and equivalent
equations (\ref{eq7}) become unjustified.

The question arises of whether there is a possibility to consistently
define the function $f_D(\tau)$ related to the frequency-dependent
permittivity (\ref{eq6}). Keeping in mind that in the case of the Drude
model the second equality in (\ref{eq5}) cannot be considered as a classical
Fourier transformation, we make an attempt to assign a definite
meaning to the function $f_D(\tau)$ by considering the generalized
inverse transformation of the quantity $\varepsilon_D(\omega)-1$
defined as
\begin{eqnarray}
f^{(0)}_D(\tau)&\equiv & -\frac{\omega_p^2}{2\pi}\int_{-\infty}^{\infty}
d\omega\,\frac{1}{(\omega+{\rm i}\,0)(\omega+{\rm i}\,\gamma)}\,
{\rm e}^{-{\rm i}\,\omega\tau}
\label{eq10} \\
&=&-\frac{\omega_p^2}{2\pi}\int_{-\infty}^{\infty}
d\omega\,\frac{\omega-{\rm i}\,\gamma}{(\omega+
{\rm i}\,0)(\omega^2+\gamma^2)}\,
{\rm e}^{-{\rm i}\,\omega\tau}
\equiv I_1+I_2.
\nonumber
\end{eqnarray}
\noindent
Here, the addition of an infinitesimally small quantity $+{\rm i}\,0$
establishes the rule on how to bypass the pole of
${\rm Im}\,\varepsilon_D(\omega)$ at $\omega=0$ and the following
notations are used
\begin{eqnarray}
I_1&=&-\frac{\omega_p^2}{2\pi}\int_{-\infty}^{\infty}
d\omega\,\frac{1}{\omega^2+\gamma^2}\,
{\rm e}^{-{\rm i}\,\omega\tau}
\label{eq11} \\
I_2&=&\frac{{\rm i}\,\omega_p^2\gamma}{2\pi}\int_{-\infty}^{\infty}
d\omega\,\frac{1}{(\omega+{\rm i}\,0)(\omega^2+\gamma^2)}\,
{\rm e}^{-{\rm i}\,\omega\tau}.
\nonumber
\end{eqnarray}

In $I_1$ the integrated function is regular at $\omega=0$.
This integral can be found in 3.354(5) \cite{6},
\begin{equation}
I_1=-\frac{\omega_p^2}{2\gamma}\left\{
\begin{array}{ll}
{\rm e}^{-\gamma\tau}\!,{\ }&\tau>0, \\
{\rm e}^{\gamma\tau}\!,{\ }&\tau<0.
\end{array}
\right.
\label{eq12}
\end{equation}
\noindent
The second integral in (\ref{eq11}) can be calculated using the
contours consisting of the real axis in the complex $\omega$-plane
and semicircles of  infinitely large radii centered at the origin in the
lower half-plane (for $\tau>0$) and in the upper
half-plane (for $\tau<0$). The result is
\begin{equation}
I_2=\frac{\omega_p^2}{2\gamma}\left\{
\begin{array}{l}
2-{\rm e}^{-\gamma\tau}\!,{\ \ }\tau>0, \\
{\rm e}^{\gamma\tau}\!,{\ \ }\tau<0,
\end{array}
\right.
\label{eq13}
\end{equation}
\noindent
where for $\tau>0$ the contributions from the two poles at
$\omega_1=-{\rm i}\,0$ and $\omega_2=-{\rm i}\,\gamma$ were taken into
account whereas for $\tau<0$  only one pole at $\omega_3={\rm i}\,\gamma$
determines the value of $I_2$. Substituting (\ref{eq12}) and
(\ref{eq13}) into the right-hand side of (\ref{eq10}) we arrive at
\begin{equation}
f^{(0)}_D(\tau)=\left\{
\begin{array}{l}
\frac{\omega_p^2}{\gamma}
\left(1-{\rm e}^{-\gamma\tau}\right),{\ \ }\tau>0, \\
0,{\ \ }\tau<0.
\end{array}
\right.
\label{eq14}
\end{equation}
\noindent
It is seen that the suggested rule leads to the same result (\ref{eq9}),
as was obtained by the inverse sine Fourier transformation from the
imaginary part of $\varepsilon_D(\omega)$ in (\ref{eq7}).

A similar situation occurs for other dielectric permittivities taking into
account free charge carriers, e.g., for the dielectric permittivities
of the plasma model and of the normal skin effect,
\begin{equation}
\varepsilon_p(\omega)=1-\frac{\omega_p^2}{\omega^2},
\qquad
\varepsilon_n(\omega)=1+{\rm i}\,\frac{4\pi\sigma_0}{\omega},
\label{eq15}
\end{equation}
\noindent
where $\sigma_0$ is the dc conductivity. In both cases the mathematical
conditions permitting to perform the Fourier transformation in the
classical understanding are violated. However, by using the same
considerations as presented above in the case of the Drude model,
one may assign a meaning  analogous to (10) to the second formula
in (\ref{eq5}) and
obtain the following dielectric permittivities as functions of $\tau$:
\begin{equation}
f_p^{(0)}(\tau)=
\left\{
\begin{array}{l}
\omega_p^2\tau,{\ \ }\tau>0, \\
0,{\ \ }\tau<0.
\end{array}\right.
\qquad
f_n^{(0)}(\tau)=
\left\{
\begin{array}{l}
4\pi\sigma_0,{\ \ }\tau>0, \\
0,{\ \ }\tau<0.
\end{array}\right.
\label{eq16}
\end{equation}
\noindent
It should be remarked that $f_p^{(0)}(\tau)$ is obtainable
also from (\ref{eq14})
in the limiting case $\gamma\to 0$.

Now let us check for consistency the respective results for
$\mbox{\boldmath$D$}(t)$. For example, we choose the electric field in the
form
\begin{equation}
\mbox{\boldmath$E$}(t)=\mbox{\boldmath$E$}_0\,
{\rm e}^{-\beta t^2}=\int_{-\infty}^{\infty}
\mbox{\boldmath$E$}(\omega)\,{\rm e}^{-{\rm i}\,\omega t}d\omega,
\label{eq17}
\end{equation}
\noindent
where $\beta>0$ and
$\mbox{\boldmath$E$}_0\equiv\mbox{\boldmath$E$}_0(\mbox{\boldmath$r$})$
describes the spatial dependence of the field. As already stated
(\ref{eq17}), the function $\mbox{\boldmath$E$}(t)$ satisfies all
required conditions and can be presented as  Fourier integral.
Its Fourier transform is calculated using the formula 3.896(4) in \cite{6},
\begin{equation}
\mbox{\boldmath$E$}(\omega)=\frac{1}{2\pi}\int_{-\infty}^{\infty}
\mbox{\boldmath$E$}(t)\,{\rm e}^{{\rm i}\omega t}dt=
\frac{1}{2\sqrt{\pi\beta}}\mbox{\boldmath$E$}_0\,
{\rm e}^{-\frac{\omega^2}{4\beta}}.
\label{eq18}
\end{equation}
\noindent
Substituting (\ref{eq14}) and (\ref{eq17}) into (\ref{eq3}) and
using the formula 3.322(2) in \cite{6} we arrive at
\begin{equation}
\hspace*{-15mm}
\mbox{\boldmath$D$}(t)=\mbox{\boldmath$E$}(t)+
\mbox{\boldmath$E$}_0\frac{\omega_p^2}{2\gamma}\sqrt{\frac{\pi}{\beta}}
\left[1+{\rm erf}\,(\sqrt{\beta}t)-
{\rm e}^{\frac{\gamma^2}{4\beta}-\gamma t}
{\rm erfc}\,\left(\frac{\gamma}{2\sqrt{\beta}}-\sqrt{\beta}t\right)\right],
\label{eq19}
\end{equation}
\noindent
where ${\rm erf}\,(x)$ is the error function and
${\rm erfc}\,(x)=1-{\rm erf}\,(x)$.
Keeping in mind that \cite{6}
\begin{equation}
{\rm erf}\,(-x)=-{\rm erf}\,(x), \qquad
{\rm erf}\,(x)=1-\frac{1}{\sqrt{\pi}}\,\frac{{\rm e}^{-x^2}}{x}+
\cdots,
\label{eq20}
\end{equation}
\noindent
we obtain for $t\to\pm\infty$
\begin{equation}
\mbox{\boldmath$D$}(\infty)=
\mbox{\boldmath$E$}_0\frac{\omega_p^2}{\gamma}\sqrt{\frac{\pi}{\beta}},
\qquad
\mbox{\boldmath$D$}(-\infty)=0.
\label{eq21}
\end{equation}
\noindent
This is what one expects on physical grounds because
for an infinite medium containing
free charge carriers the action of switching on and then
switching off electric field should result in a nonzero residual
displacement (for a finite medium, the presence of external
electric field leads to the accumulation of positive and negative
charges on the opposite boundary surfaces and to the vanishing
total electric field inside such a medium \cite{7a};
after the external electric field switches off, the accumulated
charges are distributed uniformly over the volume of
the medium leading
to zero electric displacement at $t\to +\infty$).
However, the first equality in (\ref{eq21}) means
that $\mbox{\boldmath$D$}(t)$ is not an integrable function over the
interval $(-\infty,\infty)$. This makes impossible the use of the
standard Fourier transformation (\ref{eq4}) and resulting equality
(\ref{eq5}), and makes the whole formalism not self-consistent.

This can be seen even more clearly if one defines
$\mbox{\boldmath$D$}(\omega)$ in accordance with the first equality
in (\ref{eq5}), where $\varepsilon(\omega)=\varepsilon_D(\omega)$
and $\mbox{\boldmath$E$}(\omega)$ is given in (\ref{eq18}), and then
calculates the electric displacement using the first equality in
(\ref{eq4}). The obtained quantity which we notate
$\tilde{\mbox{\boldmath$D$}}(t)$ is calculated using the formulas
3.954(2) in \cite{6} and 2.5.36(6,\,11) in \cite{7}. The result is
\begin{equation}
\tilde{\mbox{\boldmath$D$}}(t)=
\mbox{\boldmath$D$}(t)-
\mbox{\boldmath$E$}_0\frac{\omega_p^2}{2\gamma}\sqrt{\frac{\pi}{\beta}},
\label{eq22}
\end{equation}
\noindent
where ${\mbox{\boldmath$D$}}(t)$ is defined in (\ref{eq19}).
It has nonzero values at both $t\to\infty$ and $t\to -\infty$:
\begin{equation}
\tilde{\mbox{\boldmath$D$}}(\infty)=
\mbox{\boldmath$E$}_0\frac{\omega_p^2}{2\gamma}\sqrt{\frac{\pi}{\beta}},
\qquad
\tilde{\mbox{\boldmath$D$}}(-\infty)=-
\mbox{\boldmath$E$}_0\frac{\omega_p^2}{2\gamma}\sqrt{\frac{\pi}{\beta}}.
\label{eq23}
\end{equation}
\noindent
Derivation of a different electric displacement than in (\ref{eq19}), which
does not vanish at $t\to -\infty$, i.e., before the switching on of the
electric field, can be understood as an artifact resulting
from the use of the Fourier integral of a
nonintegrable function  ${\mbox{\boldmath$D$}}(\omega)$
(the definition of the Fourier integral as a generalized function
used in mathematics in this case seems to be not appropriate in our
physical situation because it is natural to understand the electric
displacement as a usual function).
This suggests that in the presence of free charge carriers the standard
definition of the frequency-dependent dielectric permittivity basing
on the formal representation of ${\mbox{\boldmath$E$}}(t)$
and ${\mbox{\boldmath$D$}}(t)$ in terms of Fourier integrals is
not satisfactory and requires some additional regularization procedure.

As an example of such procedure, we consider the modified
dielectric permittivity of the Drude model
\begin{equation}
\varepsilon_D^{(\theta)}(\omega)=1-
\frac{\omega_p^2}{(\omega+{\rm i}\theta)(\omega+{\rm i}\,\gamma)},
\label{eq23a}
\end{equation}
\noindent
where, in contrast with (\ref{eq10}), the quantity $\theta\!>\!0$ is not
infinitesimally small.
In ac\-cor\-dan\-ce with (\ref{eq23a}) $\varepsilon_D^{(\theta)}$ is
regular at $\omega=0$. The substitution of (\ref{eq23a}) into
(\ref{eq5}) leads to
\begin{eqnarray}
&&
-\omega_p^2
\frac{\omega^2-\theta\gamma}{(\omega^2+\theta^2)(\omega^2+\gamma^2)}
=\int_{0}^{\infty}
f_D^{(\theta)}(\tau)\,\cos(\omega\tau)\,d\tau,
\nonumber \\
&&
\omega_p^2(\theta+\gamma)
\frac{\omega}{(\omega^2+\theta^2)(\omega^2+\gamma^2)}=\int_{0}^{\infty}
f_D^{(\theta)}(\tau)\,\sin(\omega\tau)\,d\tau.
\label{eq23b}
\end{eqnarray}
\noindent
It can be easily seen that both the inverse cosine and sine Fourier
transformations performed in (\ref{eq23b}) lead to the common result
\begin{equation}
f^{(\theta)}_D(\tau)=\frac{\omega_p^2}{\gamma-\theta}\left(
{\rm e}^{-\theta\tau}-
{\rm e}^{-\gamma\tau}\right).
\label{eq23c}
\end{equation}
\noindent
Substituting this into (\ref{eq3}) and performing calculations with
the electric field (\ref{eq17}), we arrive at the modified electric
displacement
\begin{eqnarray}
\mbox{\boldmath$D$}^{(\theta)}(t)&=&\mbox{\boldmath$E$}(t)+
\mbox{\boldmath$E$}_0\frac{\omega_p^2}{2(\gamma-\theta)}
\sqrt{\frac{\pi}{\beta}}
\left[{\rm e}^{\frac{\theta^2}{4\beta}-\theta t}
{\rm erfc}\,\left(\frac{\theta}{2\sqrt{\beta}}-\sqrt{\beta}t\right)\right.
\nonumber \\
&&~~~-\left.
{\rm e}^{\frac{\gamma^2}{4\beta}-\gamma t}
{\rm
erfc}\,\left(\frac{\gamma}{2\sqrt{\beta}}-\sqrt{\beta}t\right)\right].
\label{eq24d}
\end{eqnarray}
\noindent
In the limiting case $\theta\to 0$ (\ref{eq24d}) coincides with (\ref{eq19}).

Precisely the same result, as in (\ref{eq24d}), is obtained if one
considers
$\mbox{\boldmath$D$}^{(\theta)}(\omega)=\varepsilon_D^{(\theta)}(\omega)
\mbox{\boldmath$E$}(\omega)$ and then finds
$\mbox{\boldmath$D$}^{(\theta)}(t)$ from the first equality in (\ref{eq4}).
Thus, when we assume $\theta>0$, both methods of the calculation
of the electric displacement are in agreement. The reason is that
for $\theta>0$ the functions $\mbox{\boldmath$D$}(t)$ and
$\mbox{\boldmath$D$}(\omega)$ belong to $L^{1}(-\infty,\infty)$ and
all Fourier transformations are well defined. However, to obtain
the correct physical results
for an infinite medium, one must put $\theta=0$ in (\ref{eq24d})
and return  to (\ref{eq19}). The point is that (\ref{eq24d}) with
$\theta>0$ leads to $\mbox{\boldmath$D$}^{(\theta)}(t)\to 0$ when
$t\to\pm\infty$ [as it must be for functions belonging to
$L^{1}(-\infty,\infty)$]. At the same time, in the presence of free
charge carriers, the electric displacement
in an infinite medium remains nonzero in
accordance with (\ref{eq21}) after the electric field is switched off.
Thus, the limiting transitions
$t\to\pm\infty$ and $\theta\to 0$ are not interchangeable.

\section{Insulating media}

The situation is quite different for dielectric
materials at zero temperature which do not contain free charge
carriers (i.e., for true insulators). In this case the dielectric
permittivity can be represented in the form \cite{8}
\begin{equation}
\varepsilon_I(\omega)=1+\sum\limits_{j=1}^{K}
\frac{g_j}{\omega_j^2-\omega^2-{\rm i}\,\gamma_j\omega},
\label{eq24}
\end{equation}
\noindent
where $\omega_j\neq 0$ are the oscillator frequencies, $\gamma_j$ are
the relaxation parameters, and $g_j$ are the oscillator strengths of $K$
oscillators.
In this case the second equality of (\ref{eq5}) results in
\begin{eqnarray}
\sum\limits_{j=1}^{K}
\frac{g_j(\omega_j^2-\omega^2)}{(\omega_j^2-\omega^2)^2+\gamma_j^2\omega^2}
=\int_{0}^{\infty}f_I(\tau)\,\cos\,(\omega\tau)d\tau,
\label{eq25} \\
\sum\limits_{j=1}^{K}
\frac{g_j\gamma_j\omega}{(\omega_j^2-\omega^2)^2+\gamma_j^2\omega^2}
=\int_{0}^{\infty}f_I(\tau)\,\sin\,(\omega\tau)d\tau.
\nonumber
\end{eqnarray}
\noindent
Performing the inverse cosine Fourier transformation in the first
equation of (\ref{eq25}) with the help of the integrals 3.733(1,\,3) in
\cite{6} we obtain
\begin{equation}
f_I(\tau)=\sum\limits_{j=1}^{K}
\frac{g_j\,{\rm e}^{-\frac{1}{2}\gamma_j\tau}}{\sqrt{\omega_j^2-
\frac{1}{4}\gamma_j^2}}\,\sin\left(\sqrt{\omega_j^2-
\frac{1}{4}\gamma_j^2}\,\,\tau\right).
\label{eq26}
\end{equation}
\noindent
Precisely the same result is obtained by means of the inverse sine
Fourier transformation from the second equation in (\ref{eq25}) when
one uses the integral 3.733(2) in \cite{6}.  In this case all involved
Fourier integrals exist in the classical sense
with no use of regularization and
$\varepsilon_I(\omega)$ is well defined.

Substituting the electric field (\ref{eq17}) into (\ref{eq3}) and using
the integral 3.897(1) in \cite{6}, we obtain the electric
displacement in an insulating media,
\begin{equation}
\hspace*{-15mm}
\mbox{\boldmath$D$}(t)=\mbox{\boldmath$E$}(t)\left\{1+
\sqrt{\frac{\pi}{\beta}}\sum\limits_{j=1}^{K}
\frac{g_j}{\sqrt{4\omega_j^2-\gamma_j^2}}\,
{\rm Im}\left[{\rm e}^{B^2}\,{\rm erfc}(B)\right]\right\},
\label{eq27}
\end{equation}
\noindent
where
\begin{equation}
B\equiv B(t)=
\frac{\gamma_j-4\beta t -{\rm i}\,
\sqrt{4\omega_j^2-\gamma_j^2}}{4\sqrt{\beta}}.
\label{eq28}
\end{equation}
\noindent
The same result is obtained by means of the inverse Fourier
transformation from $\mbox{\boldmath$D$}(\omega)$ found using the
first equality in (\ref{eq5}), as it should be. {}From (\ref{eq27}) and
(\ref{eq20}) it can be easily seen that
$\mbox{\boldmath$D$}(t)\to 0$ when $t\to\pm\infty$,
as it should be for insulating materials, and that
both $\mbox{\boldmath$D$}(t)$ and $\mbox{\boldmath$D$}(\omega)$
belong to $L^1(-\infty,\infty)$.

\section{Conclusions and discussion}

To conclude, we have shown that the definition of the frequency-dependent
dielectric permittivity for materials containing free charge carriers
by means of Fourier transformation of the fields is not as straightforward
as in the case of insulators.
The essence of the problem is in the use of the idealization of an
infinite medium. For insulators this idealization is applicable if
the sizes of the bodies are much greater than some characteristic
parameter (e.g., the width of a gap between the bodies).
However, for media with movable free charge carriers such kind of
conditions fail. The physical situation for an infinite medium turns
out to be totally different from the case of finite bodies of any
conceivable size.
In fact for conductors
$\varepsilon(\omega)$ is a quantity
obtained through formal application of Fourier transformation in the
region where it needs additional regularization.
In spite of a great number of
successful applications (see, e.g., \cite{1,2,9}) there are delicate
cases where such procedure
leads to problems. As an example one could
mention the use of $\varepsilon_D(\omega)-1$ as a response function
in the fluctuation-dissipation theorem and related
puzzles in the theory of thermal Casimir force \cite{3,10,11}.
During the last ten years the thermal Casimir force was the
subject
of considerable discussion. It was suggested \cite{14,15}
to describe it using the Lifshitz theory combined with the Drude
model (\ref{eq6}). In the limit of large separations between the
test bodies the predictions of the Drude model approach were found
to be in agreement with classical statistical physics \cite{16,17}.
On the other hand, at short separations the predictions of this
approach were excluded experimentally, whereas the predictions
based on the use of the plasma model in (\ref{eq15}) were found
to be consistent with the data \cite{18}.
The question on how to correctly calculate the thermal Casimir
force still remains to be answered. Keeping in mind that the
Lifshitz theory is based on the fluctuation-dissipation theorem,
we would like to emphasize that the application of this theorem
with poorly defined
 response functions cannot be considered as either exact
or rigorous and might cause currently discussed problems.

\section*{Acknowledgments}
The authors are grateful to V.N.~Marachevsky for helpful discussions.
 G.L.K. and V.M.M. are
grateful to the Institute
for Theoretical Physics, Leipzig University for kind
hospitality.
This work  was supported by Deutsche Forschungsgemeinschaft,
Grant No.~GE\,696/9--1.


\end{document}